\documentclass[10pt, conference, usenames,dvipsnames,svgnames,x11names,table]{IEEEtran}
\usepackage{hyphenat} 
\PassOptionsToPackage{hyphens}{url}\usepackage{hyperref}

\usepackage{xcolor}
\usepackage{booktabs} 
\usepackage{xspace}
\usepackage{tikz}
\usetikzlibrary{patterns,calc}
\usepackage{pgf-umlsd}
\usepackage{pifont}
\usepackage{url}
\usepackage{framed}
\usepackage{tcolorbox}
\colorlet{shadecolor}{Azure2}
\usepackage{enumitem}
\usepackage{amssymb}
\usepackage{cite}
\usepackage[numbers]{natbib}
\usepackage{flushend}
\newcommand{\projName}{\textsc{Mqt-Tz}\xspace}
\newcommand{\sys}{\projName}
\newcommand{\arm}{\textsc{Arm}\xspace}
\newcommand{\optee}{\textsc{Op-Tee}\xspace}
\newcommand{\qemu}{\textsc{Qemu}\xspace}
\newcommand{\tz}{\textsc{TrustZone}\xspace}
\newcommand{\mqtt}{\textsc{Mqtt}\xspace}
\newcommand{\mosquitto}{\texttt{mosquitto}\xspace}
\hypersetup{
  colorlinks,
  linkcolor={red!50!black},
  citecolor={blue!50!black},
  urlcolor={blue!80!black}
}
\usepackage{fancyhdr}

\begin{document}
\pagestyle{fancy}
\tcbset{
    colback=Azure2,
    boxrule=0.5pt, 
    boxsep=1pt,
    left=1pt,
    right=1pt,
    top=1pt,
    bottom=1pt,
    sharp corners,
    colframe=white,
}

\title{MQT-TZ: Hardening IoT Brokers\\Using ARM TrustZone\vspace{-5pt}}
\IEEEspecialpapernotice{\small{(Practical Experience Report)}\vspace{-5pt}}

\author{\IEEEauthorblockN{Carlos Segarra\IEEEauthorrefmark{1}\IEEEauthorrefmark{2}, Ricard Delgado-Gonzalo\IEEEauthorrefmark{2}, Valerio Schiavoni\IEEEauthorrefmark{1}}
\\
\vspace{-10pt}
\IEEEauthorblockA{
\IEEEauthorrefmark{2}\textit{CSEM}, \texttt{ricard.delgado@csem.ch}\\
\IEEEauthorrefmark{1}Universit\'e de Neuch\^atel, Switzerland, \texttt{first.last@unine.ch}\\ 
}
}

\maketitle


\newboolean{showcomments}
\setboolean{showcomments}{false}
\ifthenelse{\boolean{showcomments}}
{ \newcommand{\mynote}[3]{
   \fbox{\bfseries\sffamily\scriptsize#1}
   {\small$\blacktriangleright$\textsf{\emph{\color{#3}{#2}}}$\blacktriangleleft$}}}
{ \newcommand{\mynote}[3]{}}
\newcommand{\vs}[1]{\mynote{Valerio}{#1}{blue}}
\newcommand{\cs}[1]{\mynote{Carlos}{#1}{red}}
\newcommand{\rdg}[1]{\mynote{Ricard}{#1}{green}}
\definecolor{darkgreen}{rgb}{0.3,0.5,0.3}
\definecolor{darkblue}{rgb}{0.3,0.3,0.5}
\definecolor{darkred}{rgb}{0.5,0.3,0.3}

\thispagestyle{fancy}
\begin{abstract} 
The publish-subscribe paradigm is an efficient communication scheme with strong decoupling between the nodes, that is especially fit for large-scale deployments.
It adapts natively to very dynamic settings and it is used in a diversity of real-world scenarios, including finance, smart cities, medical environments, or IoT sensors.
Several of the mentioned application scenarios require increasingly stringent security guarantees due to the sensitive nature of the exchanged messages as well as the privacy demands of the clients/stakeholders/receivers.
\mqtt is a lightweight topic-based publish-subscribe protocol popular in edge and IoT settings, a \emph{de-facto} standard widely adopted nowadays by the industry and researchers.
However, \mqtt brokers must process data in clear, hence exposing a large attack surface.
This paper presents \sys, a secure \mqtt broker leveraging \arm \tz, a trusted execution environment (TEE) commonly found even on inexpensive devices largely available on the market (such as Raspberry Pi units).
We define a mutual TLS-based handshake and a two-layer encryption for end-to-end security using the TEE as a trusted proxy.
The experimental evaluation of our fully implemented prototype with micro-, macro-benchmarks, as well as with real-world industrial workloads from a MedTech use-case, highlights several trade-offs using \tz TEE.
We report several lessons learned while building and evaluating our system.
We release \sys as open-source.
\end{abstract}

\vspace{-8pt}
\section{Introduction} \label{sec:introduction}
The Internet of Things (IoT) is an increasingly popular environment to deploy all kind of data sensors, gather the produced data, and process it. 
Examples include live heart-rate data~\cite{Segarra2019}, smart-grids~\cite{Krylovskiy15}, or infrastructure management systems~\cite{Lee16}. 
The scale of IoT deployments is expected to grow exponentially in the next decade, with each individual to own and control several connected \emph{things}~\cite{turner2014digital}.
Efficient communication between the things is hence of paramount importance.

\begin{figure}[t]
    \centering    \includegraphics[width=\linewidth]{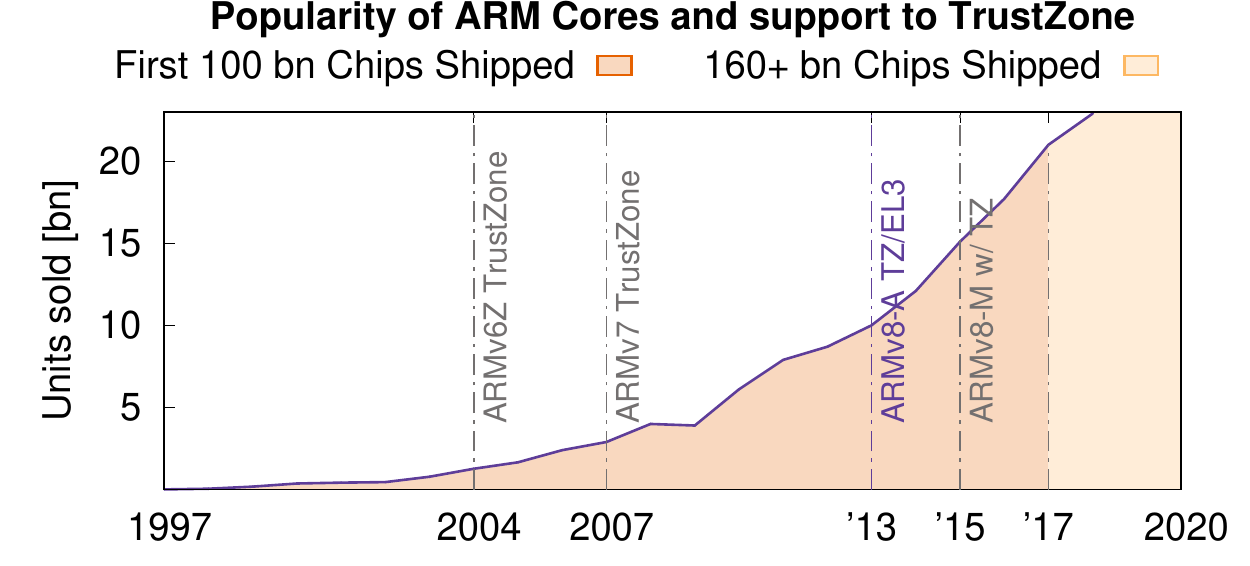}
	\vspace{-14pt}
    \caption{Sales of chips containing \arm cores since 1997. We use filled curves with the cumulative value, and highlight the years when earlier \tz versions were introduced. The current \sys can be deployed on \arm-v8-A chips, introduced in 2013.\label{fig:arm-sales}}
	\vspace{-10pt}
\end{figure}

One scalable and flexible communication pattern, commonly adopted by IoT deployers, is the publish-subscribe paradigm.
Specifically, a prominent choice in pub-sub IoT contexts is the Message Queuing Telemetry Transport (\mqtt)~\cite{mqtt-v3}, a topic-based~\cite{10.1145/1266894.1266898} publish-subscribe protocol~\cite{eugster2003many} designed for environments with limited memory and reduced network bandwidth. 
In a nutshell, a client publishes data to a \textit{topic} (\emph{e.g.}, soccer, art, \emph{etc}), while a set of \textit{brokers} forward it to the nodes subscribed to that topic.
Some deployments, including the ones that motivated our work (\S\ref{sec:use-case}) operate in sensitive environments, for which the topics, the subscriptions, or the messages being routed require high security guarantees.
Most \mqtt implementations support TLS~\cite{dierks2008transport} for transport-level security in the client-broker link, preventing malicious actors from spoofing application data. 
However, the brokers still expose a great attack surface~\cite{is-mqtt-secure}.
In particular, a powerful attacker with physical access to the broker node can intercept and tamper with all inbound and outbound traffic. 

While software-based solutions could partially mitigate the risks of such attacks, for instance exploiting homomorphic encryption (HE)~\cite{van2010fully}, the overhead that such solutions introduce in the systems is still unpractical, \textit{i.e.} nowadays still several orders of magnitude slower~\cite{8613961}. 

Trusted execution environments (TEEs) are hardware-based security mechanisms that shield code and data on compromised systems. 
Examples include Intel Software Guard Extensions (SGX)~\cite{costan2016intel}, AMD Secure Encrypted Virtualization (SEV)~\cite{kaplan2016amd} for server-grade processors, and \arm \tz~\cite{Amacher19,pinto2019demystifying,Liu2018} for edge-based processors.

\arm~\cite{arm} is a IoT leading manufacturer of IoT, from embedded devices to cloud appliances, as well as cloud-at-the-edge solutions.
It was estimated~\cite{arm-whitepaper} that by 2035 \arm will reach a trillion cumulative IoT devices shipped, with yearly sales of chips in the hundreds of billions, and trillion dollars annual spendings~\cite{arm-record-sales}.
The adoption of \arm chips in the consumer market continues to grow (Figure~\ref{fig:arm-sales}).
The support of \arm \tz across all their processors makes security become a commodity rather than an additional feature.

\tz is an edge-based TEE that enables system-wide hardware isolation against privileged processes or even malicious operating systems (\emph{i.e.}, compromised kernel or kernel modules), without the overhead of software-oriented solutions, such as the previously mentioned HE.
It is widely available on billions of consumer-grade devices (see Figure~\ref{fig:arm-sales}), most of the time for IoT applications.
For instance, the Raspberry Pi 3 Model B/B+ is a low-cost unit largely available in the consumer market.
It embeds a Cortex-A53 \arm CPU core with native support to \tz.
Additionally, \tz is easily accessible for prototyping via virtualization tools such as \qemu~\cite{Bellard2005}.
These reasons consolidate \arm \tz's position to be the ideal tool to enhance the security guarantees of existing IoT pipelines which, most likely, already run on \tz-compatible hardware.

This practical experience report presents the motivation, design, implementation, evaluation, and the lessons learned while building \projName, a secure edge-based publish/subscribe middleware leveraging \mqtt and \arm \tz protecting IoT systems against a variety of attacks.
The \sys broker exploits \tz's tamper-proof secure storage to store clients' keys (\emph{i.e.}, publishers and subscribers) upon a successful handshake phase. 
Authenticating the data publisher is not only beneficial for a trustworthy end-to-end communication, but could also be used as a digital signature when connecting a storage back-end to our secure broker.
Additionally, the re-encryption of the data---decrypting with the publisher key and encrypting with the subscriber's one---happens within the memory-protected \tz, protecting against memory-dumps. 
As detailed later (\S\ref{sec:implementation}), we built \sys's messaging broker on top of \mosquitto, the standard \emph{de facto} \mqtt implementation. 

In summary, our contributions are as follows:
\begin{itemize}[leftmargin=*]
	\vspace{-4pt}
	\item after describing our motivating scenarios from real-world settings (\S\ref{sec:use-case}), we generalize and describe the deployment scenarios (see \S\ref{sec:deployment}) for which a secure edge-based pub/sub middleware is meaningful;
    \item we describe how we protect these industrial scenarios against a powerful attacker with the available hardware, minimal additional software, and no changes to the application code running at the edge;
	\item we provide insights regarding our open-source implementation of such design. In particular, we describe a novel caching mechanism that combines \tz trusted application memory and persistent storage;
	\item we provide an in-depth evaluation of our system with real-world workloads, with the intent to highlight the performance trade-offs of \sys.
\end{itemize}
The transparency with regard to the deployer, both in terms of hardware and application compatibility, whilst adding protection for a variety of security vulnerabilities is, to the best of our knowledge, novel in the IoT/edge-processing field, and a main novelty of \sys itself.

The rest of the work is organized as follows.
We introduce some preliminary concepts on \mqtt and \tz in \S\ref{sec:background}, describe our motivating scenarios in \S\ref{sec:use-case}, and the corresponding generalized deployment settings for which \sys is most relevant in \S\ref{sec:deployment}.
The architecture and implementation details in described in \S\ref{sec:architecture}.
We present our experimental results in \S\ref{sec:evaluation}.
Finally, we cover related work (\S\ref{sec:related-work}) and our lessons learned (\S\ref{sec:lessons-learned}) before presenting our future work and concluding in \S\ref{sec:conclusion}.

\section{Background} \label{sec:background}
This section presents the technological building blocks that we leverage in \sys.
Namely, we introduce \mqtt, \mosquitto, as well as \optee, our framework and runtime of choice.
Finally, we describe \tz in a nutshell, the TEE environment available for \arm processors, to which \optee offers native access.

\textbf{\mqtt \& \mosquitto.}
The Message Queuing Telemetry Transport protocol (\mqtt) is a lightweight client and server topic-based \sloppy{publish/subscribe} messaging transport protocol~\cite{mqtt-v3,8273112}.
It is specially suited for constrained environments (\textit{e.g.}, IoT) where memory and bandwidth are very scarce resources.
In the protocol, a client publishes a message to a topic.
Dedicated processes (\emph{i.e.}, the brokers) forward it to every subscriber for that same topic.
The protocol supports decentralized deployments, in which brokers are organized in a layout and messages are routed and forwarded along the established routing tree.
The \emph{de facto} standard open-source \texttt{C}-based implementation of \mqtt is \mosquitto, actively maintained by the Eclipse Foundation~\cite{mosquitto}.
We implement \projName atop \mosquitto and introduces additional security guarantees leveraging \arm \tz, as described next.

\textbf{\tz \& \optee.}
\tz~\cite{trustzone} is a feature implemented in \arm processors since Arm1176JZ-S (2004).
It implements a trusted execution environment that is primarily used to guarantee system-wide hardware isolation.
Data and code are shielded from compromised environments.
The \tz architecture  physically separates the device in two distinct execution environments, \emph{i.e.}, the trusted side (TEE - secure world) and the untrusted side, \emph{i.e.}, Rich Execution Environment REE or normal world~\cite{Amacher19}.
\tz shields against an attacker with physical access to the device, as well as higher-privileged software or malicious kernels running in the REE.
The former hosts the so-called Trusted Applications (TAs). 
Trusted applications can leverage additional \tz-only services, such as tamper-proof persistent storage via specialized APIs.
The integrity of a \tz-enabled device can be guaranteed by a secure boot mechanism.\footnote{While beneficial, we did not leverage secure boot to build \sys since our Raspberry Pi 3 Model B devices do not support it.} 
Its root of trust builds on Hardware Unique Keys (HUK) embedded in the processor during manufacturing.
A common assumption is that only trustworthy TAs are deployed inside the secure world.
The Open Portable Trusted Execution Environment (\optee)~\cite{optee} is an open-source runtime for TEE applications sponsored by the Linaro Foundation~\cite{linaro-web} with native support for \tz.
It is designed to run together with a non-secure Linux kernel in the REE.
Finally, \optee is compliant with the GlobalPlatform's specifications~\cite{globalplatform-specifications}.

\section{Motivating Scenarios}\label{sec:use-case}

\begin{figure}[!t]
    \includegraphics[scale=0.62]{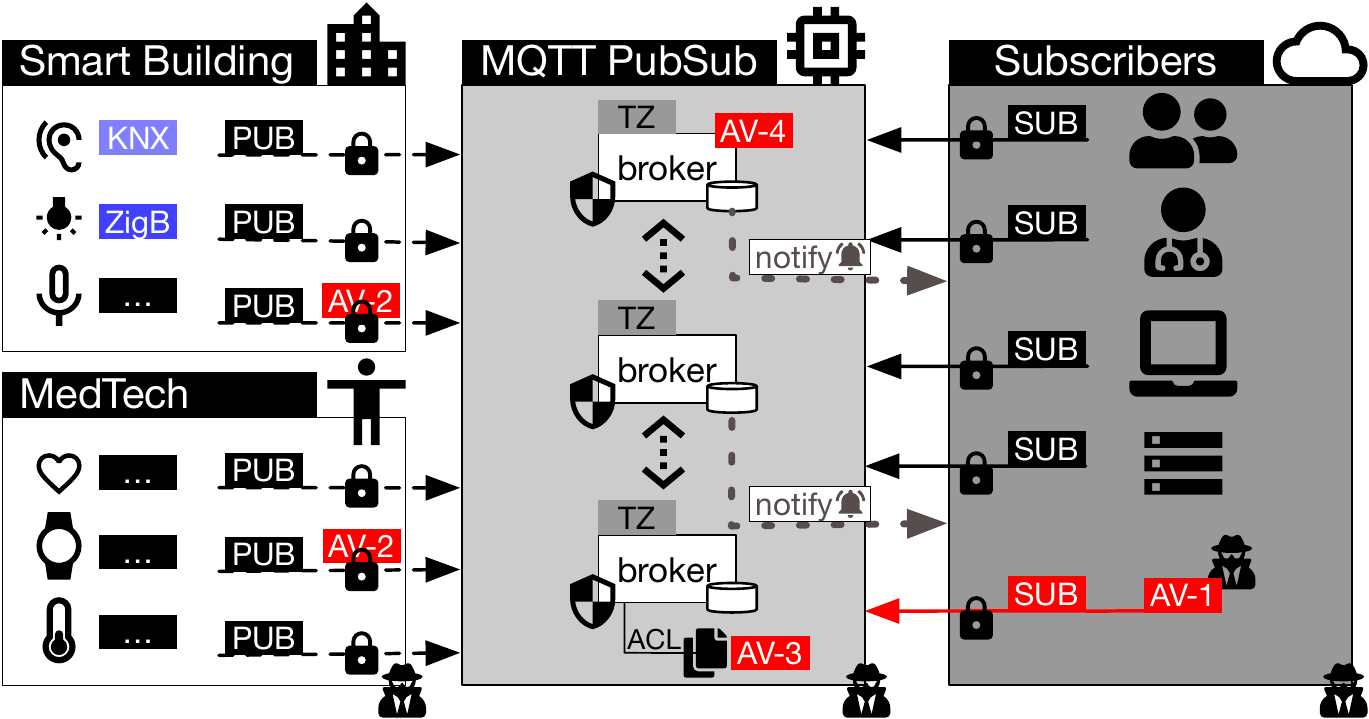}
    \caption{Our two motivating scenarios: smart building management and vital signs monitoring (MedTech). The figure shows the interaction of the various actors with the \mqtt publish-subscribe middleware as well where the identified attack vectors can be launched (\S\ref{sec:deployment}).~\label{figs:usecases}}
\end{figure}

In this section, we describe our two main real-world scenarios behind \projName, both depicted in Figure~\ref{figs:usecases}. 
The use-cases originate from two on-going research projects, in collaboration with industry-leading companies.
The security vulnerabilities exposed in the following and that we tackle in the reminder are both realistic, and believed to be currently exploitable at the moment.
In particular, the security and privacy concerns present in both cases are similar and can be attributed to:
\begin{itemize}
    \item the exposure of the \mqtt broker to attackers at the edge, both from attackers with physical access (on the field) to the devices, or by means of privileged malicious software running in the same node;
    \item the lack of authentication and authorization in the communication channel;
    \item the lack of transport and application layer protection;
\end{itemize}
\subsection{Smart Building Management}\label{sec:use-case:tabede}
The first scenario stems from TABEDE\footnote{\url{https://www.tabede.eu/}}, an EU H2020 project with the aim to integrate energy grid demand-response schemes into buildings through low-cost extenders for Building Management Systems or as a standalone system, which is independent of communication standards and integrate innovative flexibility algorithms.
The project targets the control of building mechanical and electrical equipment and services such as air conditioning, ventilation, and security systems.

From the architectural perspective, the overall system can be summarized on three key parts: \emph{(1)} several in-house smart sensors implementing a large variety of communication protocols (\textit{e.g.} EnOcean~\cite{li2014bacnet}, KNX~\cite{lee2009implementation}, Zigbee~\cite{farahani2011zigbee}); \emph{(2)} a back-end managed by a third-party (\emph{e.g.}, an untrusted cloud provider) where the data is streamed and stored in real time; and \emph{(3)} a web-based front-end that visualizes and manages the network of in-house sensors.

The flow of information relies on \mqtt brokers deployed at the edge to minimize latency and limits the physical access from untrusted entities. 
However, it directly raises several privacy and security concerns. 
In particular, the energy readings of an ensemble of buildings represent vital information required for the energy-balancing load algorithms that manage the internal power grid. 
If the readings are tampered with, even by a small amount, the grid could suffer local overloads. 
This method could be used to deliberately attack a building by overloading its grid (or even worse, if the attack is launched at a national scale~\cite{wired-hack}). 
Finally, in terms of privacy concerns, energy consumption reveals details about the habits and behaviors of individuals inside their private homes.
Therefore, smart metering has to compromise between detailed energy metering and privacy protection~\cite{McDaniel2009,Cuijpers2013}.
This use-case helps us identifying the following open questions:
\begin{tcolorbox}
    \textit{\textbf{Q1}: What if an attacker impersonates the power meter and sends erroneous power readings to the grid?} 
\end{tcolorbox}
\begin{tcolorbox}
\textit{\textbf{Q2}: What if an attacker compromise the smart-light or smart-lock communication with the \mqtt broker?} 
\end{tcolorbox}

\vspace{-12pt}
\subsection{MedTech for Vital Sings Monitoring}\label{sec:use-case:medtech}

In the context of medical technologies (MedTech), the monitoring of vital signs is increasingly off-loaded and outsourced to third-party untrusted data centers.
The main reason is to exploit the economy of scale that comes with cloud computing. 
However, recent data protection regulations (\textit{e.g.}, GDPR~\cite{voigt2017eu}) have stressed the importance of ownership of the data and limited the scope of its use.
Much research has been recently devoted to deal with such restrictions and on how to reconcile distributed systems with such legislation frameworks~\cite{berberich2016blockchain}. 
For instance, \cite{Segarra2019b} showed that it is possible to deploy real time-processing of heart-rate variability (HRV, linked to physiological and mental stress~\cite{Tan2011,Castaldo2015}) by exploiting Intel SGX TEE enclaves with reasonable overheads.
However, the platform must deal with a large diversity of physiological and behavioral signals originating from wearable sensors~\cite{DelgadoGonzalo2017,Renevey2017} such as electrocardiograms (ECG), electroencephalograms (EEG), or photoplethysmographams (PPG).

In this scenario, the flow of information is typically mediated by \mqtt brokers. 
More precisely, ECGs and other key signal features are sent to remote \mqtt brokers for further processing like HRV monitoring and arrhythmia detection. 
Locating them at the edge of the network, rather than the cloud, guarantees that data ownership is not transferred, hence easing compliance with data protection regulations.
If a health anomaly is detected, an alarm is relayed through an \mqtt topic to which a particular doctor or emergency services is subscribed. 
This architecture raises similar privacy and security concerns as the smart building scenario (\S\ref{sec:use-case:tabede}) since, vital signs, and in particular HRV signals, are very sensitive data as well.

Moreover, as our healthcare systems quickly transition towards personalized medicine~\cite{Hamburg2010}, an erroneous diagnosis generated by tampered health data could result in life-threatening situations. 
On the privacy side, read access to this health data can be used by third parties for user profiling, which may be used by health insurance companies to raise premiums from the leaked information.
\begin{tcolorbox}
\textit{\textbf{Q3}: What if an attacker deliberately push data to one patient's HRV topic, invalidating all the further monitoring?} 
\end{tcolorbox}
\begin{tcolorbox}
\textit{\textbf{Q4}: What if an attacker compromise the medical broker to tamper packets routed to the emergency topic?} 
\end{tcolorbox}

\vspace{-10pt}
\section{Deployment Scenario \& Threat Model} \label{sec:deployment}
From the described use-cases, we characterize our deployment scenarios used to design \sys.
This section also describes the targeted threat model, as well as to clearly identify various attack vectors (\emph{AV}) form which \sys shields against.

\textbf{Deployment Scenario.}
We envision a deployment scenario consisted of a large set of low-powered, memory-constrained client and server nodes.

Client nodes continuously publish live monitoring data in a streaming fashion.
These nodes can typically sustain a steady network throughput of hundreds of bytes per second. For instance, a publisher streaming an ECG will send at most 350 Bytes/s, and a full fleet of publishers (\emph{e.g.}, a floor in a hospital) will amount to around 3-5 kBytes per second (see \S\ref{sec:evaluation}).

Server nodes in the system are placed at the edge. 
This choice maximizes responsiveness (minimizing latency), reduces the attack surface, and avoids the transferring of data ownership.
They receive such data to process, for instance performing aggregation, averaging, or detecting statistical deviations.
Increasingly common in IoT deployments---given its reduced costs, high availability in the market, popularity across developers, and hardware features---we deploy our brokers over Raspberry Pi 3 Model B units (see \S\ref{sec:medtech-in-action}).
Conveniently, this device embeds an \arm Cortex-A processor with native support for \tz. 
The deployment scenario includes storage services in charge of collecting data to be processed offline at a later time.

\textbf{Threat Model.}
Figure~\ref{figs:usecases} depicts the potential threats that we consider in our deployment scenario, including the several attack vectors at disposal of an attacker.
Note that these are not flaws in the \mqtt protocol \emph{per se} as they may be out of scope (\emph{e.g.}, transport layer security).
We highlight flaws in the way these tools and protocols are used in industrial settings.

First, most IoT settings leveraging \mqtt bypass client or broker authentication (\emph{AV-1}).
As a consequence, unauthorized parties can publish data to the brokers as long as their address is known. 
Similarly, attackers might try to impersonate brokers. 

Second, they do not encrypt data packets before transmission.
As a consequence, an attacker with the ability to intercept or spoof the client-broker link would gain access and tamper all the information processed by the broker (\emph{AV-2}).

Third, edge-based pub/sub middleware usually lacks any mechanism of access control, \emph{e.g.}, the topics are public. 
An attacker with knowledge of the publish/subscribe topics could inject carefully crafted information while receiving potentially sensitive data sent by the clients (\emph{AV-3}).
By leveraging access control mechanisms, we can also revoke access to byzantine nodes, hence providing a simple defense mechanism against replay attacks towards the broker.

Lastly, the \mosquitto \mqtt brokers run by default as non-privileged processes.
This exposes brokers to higher privileged processes running in the same node (\emph{e.g.}, in the form of malware as these brokers tend to be connected to the internet), malicious users (\emph{e.g.}, with SSH access to the node), or even malicious operating systems loaded by an advanced attacker at boot time.
For instance, a high-privilege process running on the same broker machine could intercept all its incoming and outgoing traffic, tampering with the data it processes.
Note also that even when the client-broker link is encrypted  (\textit{e.g.}, via TLS channels), processing data in clear at the broker constitutes a privacy and security risk (\emph{AV-4}).
To protect against an attacker patching our broker, or rebooting with a different implementation which could trick the untrusted code to authenticate malicious clients, we measure the REE code base at secure-boot time.

When compared to lightweight cryptographic-based schemes implementing device authentication and access control mechanisms, \sys offers the additional in-broker security guarantees which stem from the usage \tz.

We consider a powerful attacker with administrative rights as well as physical access to the broker node, and with the ability to exploit any of the listed attack vectors.
However, we assume the client to not be compromised.
While we are aware of recent \tz side-channel attacks~\cite{10.1145/3319535.3354197,10.1145/3319535.3354201}, we consider those out-of-the-scope of this work.
In particular, we assume that the client's cryptographic credentials (see further details in \S\ref{sec:architecture}) cannot be extracted. 

\section{The \sys System} \label{sec:architecture}
This section presents the architecture and the interaction of the various \sys components, providing additional implementation details, as well as explaining how the open questions given in \S\ref{sec:use-case} are addressed.

\begin{figure}[t!]
    \centering
    \includegraphics[scale=0.65]{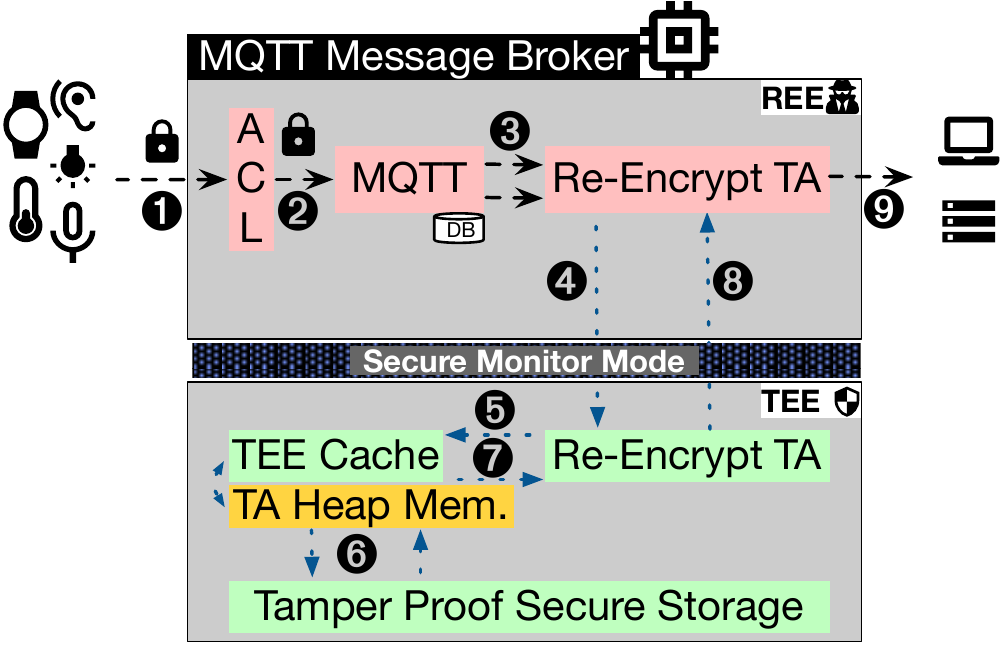}
    \caption{\projName architecture and flow of operations.\label{fig:architecture}}
	\vspace{-20pt}
\end{figure}

\subsection{Architecture \& Component Description}
The architecture of \projName is presented in Figure~\ref{fig:architecture}.
We extend the \texttt{mosquitto} \mqtt broker to \arm \tz.
Specifically, \sys's broker adds an encryption layer in \mqtt's payload using client-specific keys stored inside \arm's secure storage~\cite{optee-secure-storage}.
Doing so, application data is only processed inside the TEE, where it gets re-encrypted.
For the additional key-provisioning to address AV1 and AV2, we redefine the client authentication in the mutual TLS handshake to prevent the REE from gaining access to clients' keys.
We also leverage Access Control Lists (ACLs) for fine-grained control over the users and their topics.
We now detail the various architecture components and their mutual interaction.

Note that \sys does not modify nor redefine TLS or the vanilla \mqtt protocol.
We leverage---and modify when needed---existing implementations of these protocols to shield our deployment scenario from possible attacks (\S\ref{sec:deployment}), with the minimum performance penalty, as shown next.

\begin{figure}[t!]
    \centering
    \includegraphics[width=\linewidth]{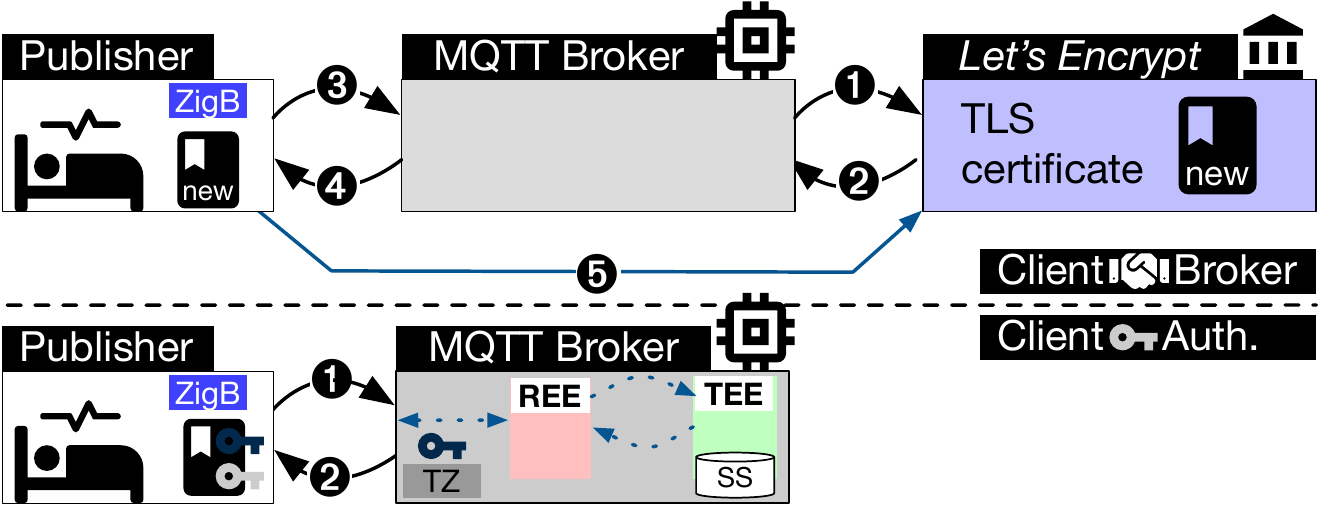}
    \caption{\sys: broker (above) and client (below) authentication.\label{fig:handshake}}
	\vspace{-10pt}
\end{figure}

\textbf{Two-Step Handshake.}
\projName defines and uses a two-step handshake that realizes broker and client authentication with end-to-end encryption from the client to the TEE.
The handshake protocol requires minimal pre-provisioned cryptographic material.
\begin{tcolorbox}
\textit{\textbf{A1}: mutual authentication prevents from either broker or client impersonation.}
\end{tcolorbox}
\vspace{-5pt}
The broker authentication (Figure~\ref{fig:handshake}, top) relies on TLS handshake, supported by default in \texttt{mosquitto}, while the corresponding client step (Figure~\ref{fig:handshake}, bottom) uses \mqtt. 
We choose the latter over TLS's client side authentication to ensure that client's data is only processed in clear inside the TEE.
Alternatively, we would need to install a full TLS endpoint inside \tz, leading to a very large attack surface as code in the TEE \emph{needs} to be trusted.

First, the client publishes its symmetric key, encrypted with the broker's TEE public key, to a dedicated write-only topic.
This TEE key pair is generated at device start-up time (secure boot) and derived from a Hardware Unique Key (HUK).
As a consequence, the private key never leaves the \tz.
Note that secure boot and HUKs are device-specific, hence the exact mechanism depends on the system on which \projName is deployed. %
The encrypted payload is then securely transferred to the \tz TEE and decrypted.
The client's key is stored in the secure storage (\texttt{SS} in Figure~\ref{fig:handshake}, bottom).
An \texttt{ACK} reply is encrypted with this same symmetric key and sent back to the REE, which can forward to the client to finalize the handshake.

\begin{tcolorbox}
    \textit{\textbf{A2}: TLS protects the communication link and prevents malicious packet interception and man-in-the-middle attacks.}
\end{tcolorbox}
\vspace{-5pt}
\textbf{Layered Encryption \& Access Control Mechanisms.}
After the handshake, \projName uses a two-layer encryption mechanism.
First, the client-broker link is protected by TLS within \mosquitto.
Second, \mqtt's payload is encrypted using the clients' symmetric key. 
Then, data is re-encrypted in the TEE (as detailed next) and sent to its destination over a \mosquitto-TLS channel.
Doing so, we achieve end-to-end security relying on \tz as a secure proxy.
We also leverage Access Control Lists to limit the clients able to interact with the broker.
These are currently stored in the REE, but if clients are defined in advance, its contents could be measured during boot and stored also in \texttt{SS}, preventing an attacker from tampering with the lists.
\begin{tcolorbox}
    \textit{\textbf{A3}: ACLs prevent unauthorized entities to inject or subscribe to sensitive topics, and enable revoking access to clients controlled by attackers.}
\end{tcolorbox}
\vspace{-5pt}
\textbf{Payload Re-encryption.}
We link \mosquitto with a trusted application that transfers the encrypted data to \tz's secure world, where it retrieves the origin and destination keys from secure storage, and re-encrypts the information.
Currently, topic subscription lists and \mqtt metadata are stored in a dedicated database (\mqtt DB) in the REE. 
We plan on shadowing these structures to keep them in the TEE since information like subscription patterns, subscriber distribution, or topic filtering can be used as a side-channel to leak sensitive data.
\begin{tcolorbox}
    \textit{\textbf{A4}: The re-encryption of information \textbf{inside} \tz prevents a physical attacker or privileged process from spoofing sensitive information.}
\end{tcolorbox}
\vspace{-5pt}
\textbf{LRU Cache in the TEE.}
Our evaluation (\S\ref{sec:evaluation}) shows that retrieving the keys from secure storage is the most expensive operation when re-encrypting application data in the TEE.
To mitigate this behavior, \sys embeds a lightweight LRU cache in the TEE that keeps the most recently used keys in the TA's heap memory, and evicts the least used ones (LRU policy) to persistent secure storage.
Access to the heap's content is also hardware-protected by \tz.

\textbf{Dataflow.} In a nutshell, data in \sys flows as follows.  
Data travels two-fold encrypted from the client to the broker (Fig.\ref{fig:architecture}-\ding{202}). 
Once the client access is confirmed (Fig.\ref{fig:architecture}-\ding{203}), the subscribers for the given topic are retrieved and the payload forwarded (Fig.\ref{fig:architecture}-\ding{204}).
Then, encrypted data is transferred to the TEE (Fig.\ref{fig:architecture}-\ding{205}).
The origin and destination client keys are retrieved (\ding{206}-\ding{208}).
The payload is re-encrypted and sent back to the REE (Fig.\ref{fig:architecture}-\ding{209}) and to the subscriber (Fig.\ref{fig:architecture}-\ding{210}).

\begin{figure*}[t!]
    \centering
    \includegraphics[width=\linewidth]{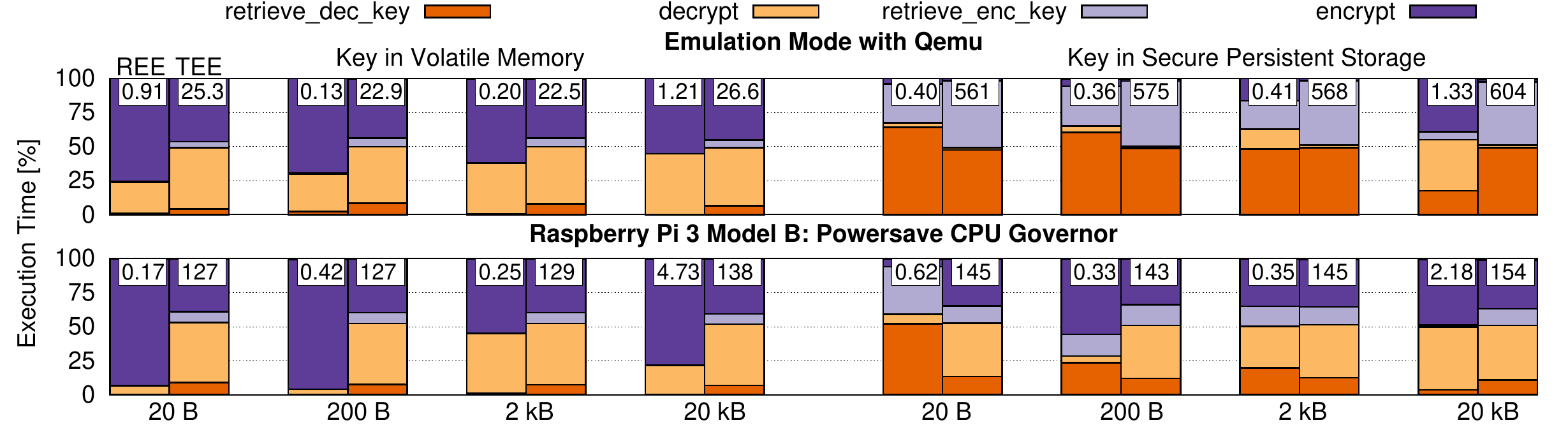}
    \caption{Re-encryption TA microbenchmark. Percentage breakdown of the contributions of each operation in the time elapsed. For each different block size (bottom axis) we compare: re-encryption in the REE vs TEE with the key stored in memory or persistent storage. We also report average total time (in ms) boxed. \label{fig:ub1}}
\end{figure*}

\subsection{Implementation Details} \label{sec:implementation}

\projName's broker is implemented in \texttt{C}.
The current version of \projName adds 400 SLOC to \texttt{mosquitto} version 1.6.3 and the TA amounts to 1204 SLOC. 
The \sys TA relies on \optee, version 3.5.0. 
The \sys prototype will be available from \url{https://github.com/mqttz}.


\textbf{Client and Server Authentication.}
The server-side authentication is done through vanilla TLS.
We deploy \projName's secure broker in a device with a static IP address. 
Then, we bound the address to a domain name and use a certificate.
We rely on \textit{Let's Encrypt} (\url{https://letsencrypt.org/}) to get one and to authenticate the broker.
The client-side authentication uses \mqtt as communication layer, and \texttt{openssl} (v1.1.1a) for cryptographic tools.
The integration with \texttt{mosquitto} exploits custom callbacks for each packet processing.
In addition, we use \mqtt (v5.0) Request/Response (RR) features for the client's key exchange.
To control access and R/W permissions to topics, we use \texttt{mosquitto}'s ACLs.

\textbf{Trusted Application.}
We rely on \optee APIs to implement the payload re-encryption TA.
Trusted applications implemented within this framework have two parts: \emph{(1)} a host app that runs in the REE and acts as an entry point and bridge to the TEE, and \emph{(2)} a trusted API in the TEE that exposes different functions. 
\projName intercepts all \mqtt packets forwarded to the recipients, and feeds our host app with both client's IDs and the encrypted data.
Then, the payload re-encryption happens, using \optee's storage and cryptographic libraries.
\tz not only provides isolation between worlds, but also between different TAs.
Hence, we use the same secure API to store new keys during the handshake.
For key retrieval, we implement a new LRU cache in the TEE to store the most frequently used keys in the TA's heap, while the remaining ones are evicted and flushed in persistent secure storage~\cite{optee-secure-storage}. 

\section{Evaluation} \label{sec:evaluation}

We present the experimental evaluation of the \sys prototype using micro-benchmarks and macro-benchmarks, as well as using real-world datasets from the MedTech scenario.
Our intent is to validate the design of \sys, the efficiency of our implementation and to analyze the different trade-offs that the system incurs.

\textbf{Evaluation Settings.}
We use Raspberry Pi 3 Model B units, one of the few where \optee fully supports \tz.\footnote{\url{https://optee.readthedocs.io/en/latest/building/devices/rpi3.html\#what-versions-of-raspberry-pi-will-work}}
We set up the CPU in \texttt{powersave} governor mode to minimize energy consumption.
   
To validate the scalability of \sys, we also deploy a fleet of virtualized nodes using \textsc{Qemu-v8}\footnote{\url{https://www.qemu.org/}}, as it faithfully replicates the industrial settings planned for \sys. 
This emulated environment closely matches the expected hardware performance~\cite{Amacher19}. 
We use \texttt{mosquitto} (v1.6.3) and \optee (v3.5.0).
Unless otherwise specified, messages are 4 kBytes and are encrypted with 32-Byte keys.
We always report average and standard deviation results for 100 executions of the described configurations. 
The default size of the messages reflects the setting presented in \S\ref{sec:medtech-in-action}.
\begin{figure}[t!]
    \centering
    \includegraphics[scale=0.68]{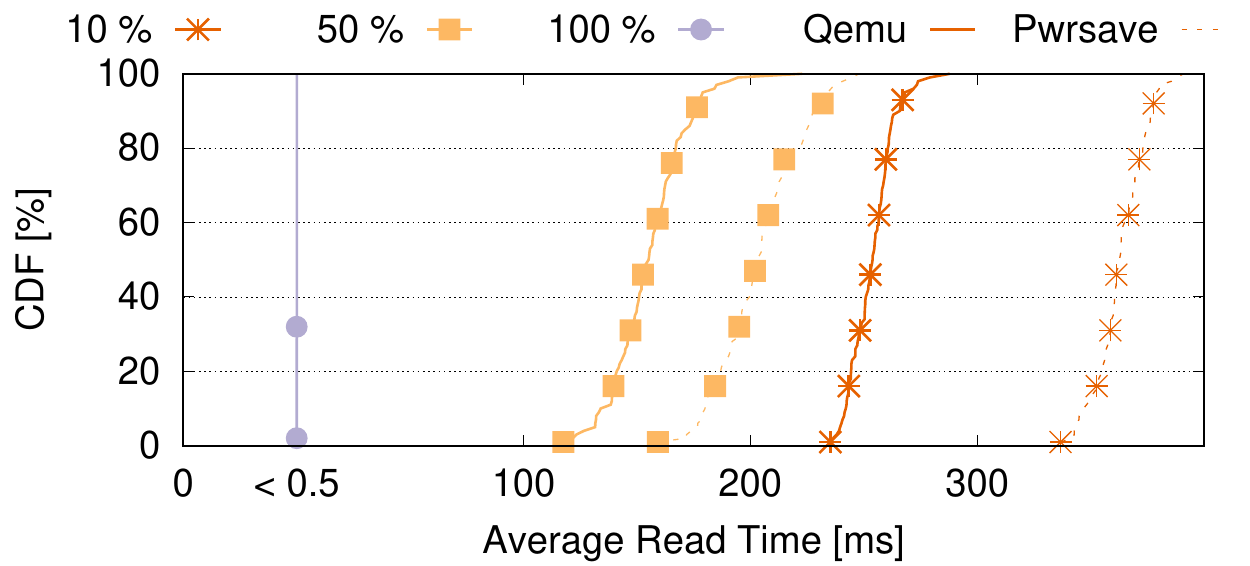}
    \caption{Read time (CDF) from secure storage using \sys's secure cache for different numbers of cached objects. The cache capacity grows from 10\% up to 100\% of the total objects. Dashed lines represent hardware results and bold lines emulated ones.\label{fig:cache-benchmark}}
\end{figure}

\subsection{Micro-benchmarks}
\textbf{Re-Encryption TA.}
We begin by measuring the time required to re-encrypt a block of data inside or outside the TEE, one of \sys's cornerstone operations.
We include results for the hardware and the emulated environment using the \texttt{powersave} CPU governor, as it most closely matches the expected deployment settings.
We breakdown these measures into four main components: the time it takes to retrieve each key (\texttt{retrieve\_dec\_key}, \texttt{retriev\_enc\_key}), and to use them (\texttt{encrypt}, \texttt{decrypt}).
On the x-axis, we show results for different block sizes of data to re-encrypt, \emph{e.g.}, from 20 Bytes up to 20 kBytes.
We compare the performance of \sys decrypting inside the REE (left-side vertical bars) or in the TEE (right-side bars). 
For encryption, we use AES in CBC mode with 32-Byte keys.
Finally, we include results for the variants of the system that maintain the keys either in volatile memory as well as in secure storage, and the overall elapsed time (in ms) boxed.
Figure~\ref{fig:ub1} uses a stacked bar chart representation to present these results.

We observe that AES symmetric cryptography is two orders of magnitude slower in the TEE, both in emulation and in real hardware.
For instance, for the 2kB case, we face a slowdown of 128$\times$ and 250$\times$, respectively.
This is due to the cryptographic libraries in \optee not using hardware accelerators (due to security restrictions) and hence not as optimized as \texttt{openssl} used in the REE. 
Moreover, when switching from in-memory to secure persistent storage, we observe even higher slowdowns (\emph{i.e.}, up to 250$\times$ and 260$\times$ in the 20 Bytes case), specifically in the time required to fetch the decryption key.
These results motivate the inclusion of a LRU cache in \projName's architecture, as described next.

The emulation results show slower memory access times, but faster encryption both in the TEE and REE when compared to real hardware.
In spite of that, emulation closely matches the time spent in each phase specially for the \textit{Key in Volatile Memory Case}.
When the key is stored using the Secure Storage API, \textsc{Qemu} overestimates the time to retrieve both objects, and underestimates the time to encrypt them.
Therefore, emulation is useful for prototyping and early stages of development, but to draw conclusions on performance one must measure on the real hardware, under realistic workloads.
\begin{figure}[t!]
    \centering
    \includegraphics[scale=0.68]{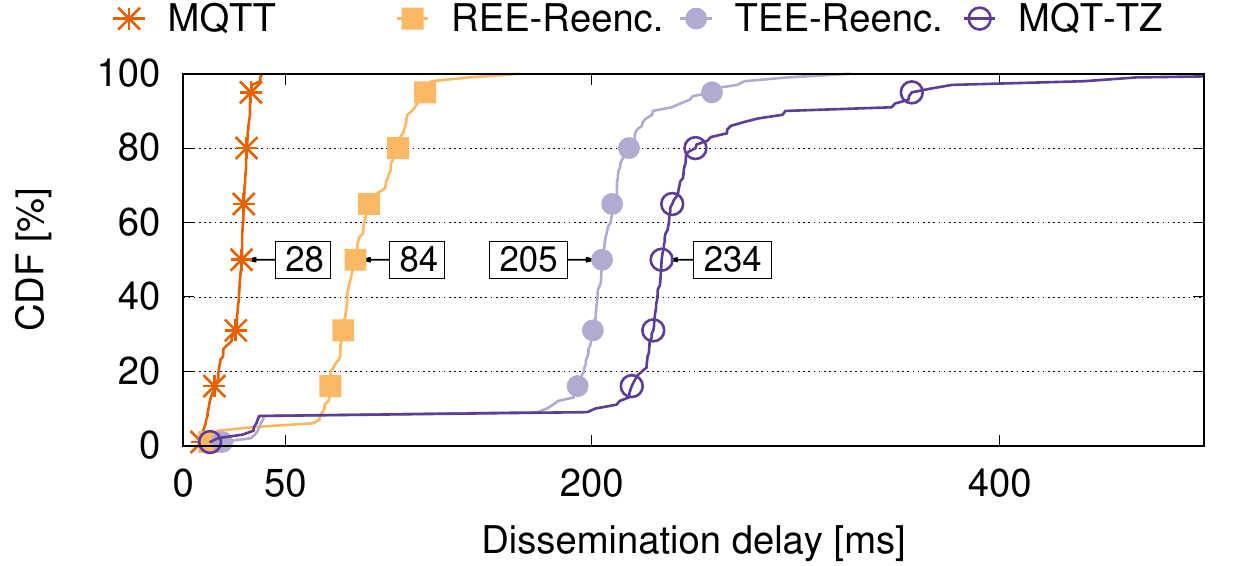}
    \caption{Dissemination delays (CDF) for different \mqtt implementations and configurations of \sys. We highlight (boxed) the delay in ms for the 50th percentile. \label{fig:mb1}}
\end{figure}
\textbf{TEE Cache.} To evaluate the performance of our LRU cache inside \tz, we issue \texttt{get} queries to fetch entries from the cache. 
Figure~\ref{fig:cache-benchmark} presents the cumulative distribution function (CDF) of the delays to return and read the reply.
To execute this experiment, we initially preload a set of 128 256-bits AES keys to secure storage, as this matches the expected size of the system (both in terms of subscribers and publishers) that \sys will support in real-world deployments. 
Then, the cache is filled with keys randomly sampled from this set.
The number of entries in the cache varies between 12, 64, and 128 keys, \emph{i.e.},  10\%, 50\%, and 100\% of the total.
The client issues 128 random queries, and measure the average latency over 100 runs for each configuration.
Note that clock precision for such in-TEE measurements is 1~ms.\footnote{This is the default precision of the \texttt{TEE\_Time}~\cite{tee-time} offered by \optee. 
} 
Hence, replies faster than this threshold (\emph{e.g.}, as in the 100\% case) are not reported.
As expected, smaller caches lead to more the cache misses and higher average reading time.
The median (50\textsuperscript{th} percentile) reading value for the 50\% case is 155~ms, and up to 253.85~ms for the 10\% scenario.
These results indicate the LRU cache is beneficial in all the tested scenarios. 

When comparing hardware (dashed) with emulated (bold) values, we observe that \textsc{Qemu} fails to emulate the object storage and retrieval performance when the secure store is relatively full (128 keys in our case).
This is specially true when using the \texttt{powersave} CPU governor mode~\cite{Amacher19}.

\subsection{Macro-benchmarks}
To assess the overall performance of \sys, we measure the dissemination delay (latency) for a \sys setup with one single publisher and one single subscriber.
We additionally show a larger-scale deployment scaling up the number of subscribers to the same topic.
These benchmarks stress the operations occurring the most inside \tz, \emph{i.e.} encryption, decryption, and queries to the cache.
Increasing the number of publishers (instead of subscribers) would yield the same number of cache queries, hence a symmetric effect on performance.

\textbf{1 Publisher - 1 Subscriber.}
This is a baseline deployment, used to assess the dissemination delay performance.
We compare vanilla \mosquitto against several variants of \sys: \emph{(1)} with re-encryption in the REE and all keys in memory, \emph{(2)} re-encryption in the TEE and all keys in memory, and \emph{(3)} all the features combined.
Figure~\ref{fig:mb1} reports the CDF of the dissemination delays.
As expected, an increasing number of security features hurts the dissemination delays (up to 8$\times$ for the median values), with a long tail up to 350~ms. 

\begin{figure}[t!]
    \centering   
	\includegraphics[scale=0.68]{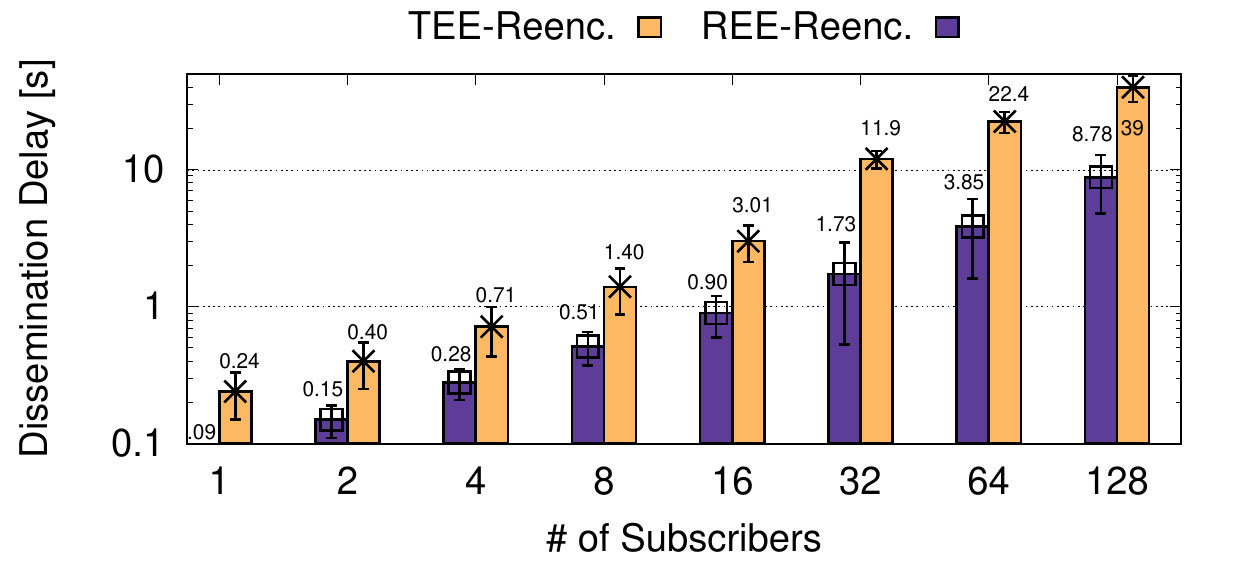}
    \caption{Impact of total subscribers per topic on the overall message dissemination delays.\label{fig:sub-scal}}
\end{figure}

\textbf{1 Publisher - Many Subscribers.} Next, we scale up the number of subscribers for a given topic.
Figure~\ref{fig:sub-scal} presents the dissemination delays for two configurations (REE-Reenc. and TEE-Reenc.) \emph{e.g.}, using re-encryption in the REE and TEE respectively.
We observe that the dissemination delays increase linearly with the number of subscribers.
Re-encryption in the TEE (in general TA execution) being single-threaded, each subscriber has to be handled individually.
As a conclusion, the biggest performance bottlenecks derive from the slow cryptographic primitives and the key retrieval routines.
For the former, we rely on the primitives provided by \optee.
We intend to exploit Cryptographic Accelerators for \tz (\emph{e.g.} \arm \textsc{CryptoIsland}~\cite{arm-cryptoisland}) to considerably speed-up these operations.
For the latter, we introduced an in-TEE cache to minimize the queries to persistent storage.
Another strategy along this line would be to share one key among sets of clients (if the application's security concerns allowed to).

In terms of scalability, \sys is limited by the per-subscriber re-encryption process.
This process being single-threaded (\optee's limitation), all clients are serialized.
A possible workaround would be to leverage \mosquitto's multithreading capacities combined with multiple running concurrent TAs (not in-TA threading).
However, this would incur in higher CPU load and hence higher energy consumption, a possible showstopper in some constrained environments.
Alternatively, establishing per-topic shared keys among client nodes, the re-encryption for some packets could be avoided. 
We intend to explore these options in future extensions of this work.
\vspace{-6pt}
\subsection{MedTech in action} \label{sec:medtech-in-action}
\begin{figure}[t!]
    \centering
    \includegraphics[scale=0.68]{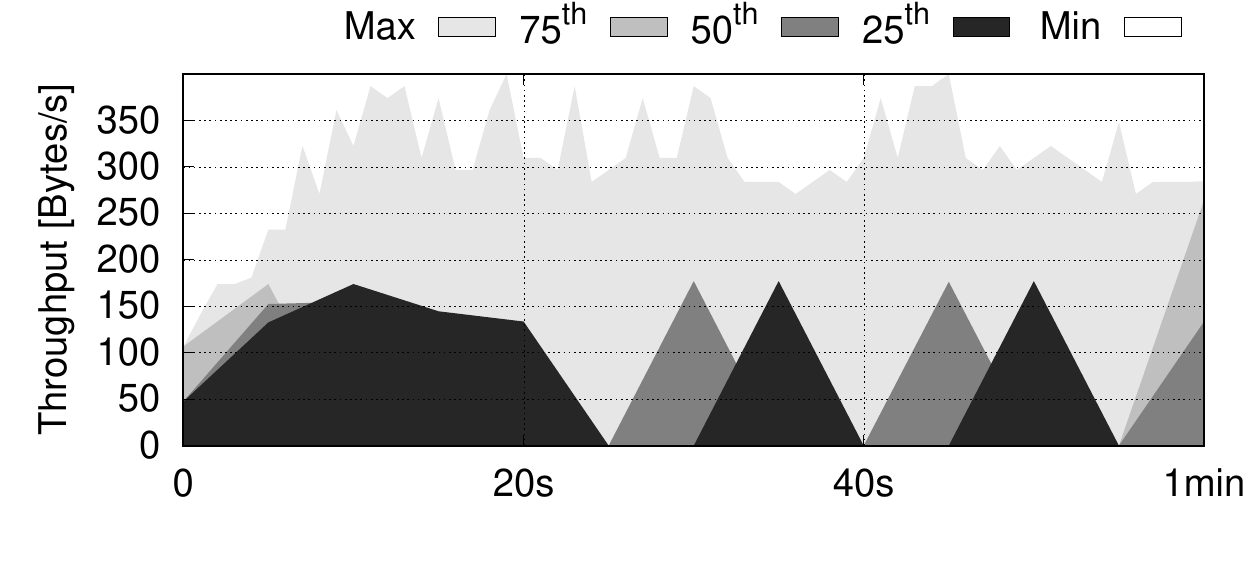}
    \caption{Distribution of network throughput for a realistic deployment with 50 publishers emitting real-world cardiac signals.\label{fig:out-traffic}}
\end{figure}

\vspace{-2pt}
Finally, we fully implement and deploy the vital signs monitoring scenario (\S\ref{sec:use-case:medtech}).
In this case, we are interested in understanding if in a real-world setting the \sys broker can efficiently (\emph{e.g.}, CPU processing) sustain the injected workload.
We leverage real-world ECG datasets we collected on the field. 
This deployment reproduces the layout of an hospital floor with 50 patients whose cardiac signals are constantly monitored.
For the sake of simplicity, these signals are streamed toward one single \sys broker, although a federated deployment is also supported.
We capture and measure the outbound network traffic from each publisher using \texttt{nethogs}~\cite{nethogs}.
Figure~\ref{fig:out-traffic} depicts the outbound throughput generated by each publisher in bytes per second.
We use a stacked percentile representation with shades of grey to plot the minimum, 25\textsuperscript{th}, the 50\textsuperscript{th} (median), the 75\textsuperscript{th} and the maximum across all the publishers.
We observe that at any given time only a subset of the publishers actually emits data.
A single subscriber streams at 350 Bytes/s in the worst case, and the full fleet of publishers generates between 3 to 5 kBytes per second overall.

\begin{figure}[t!]
    \centering
    \includegraphics[scale=0.68]{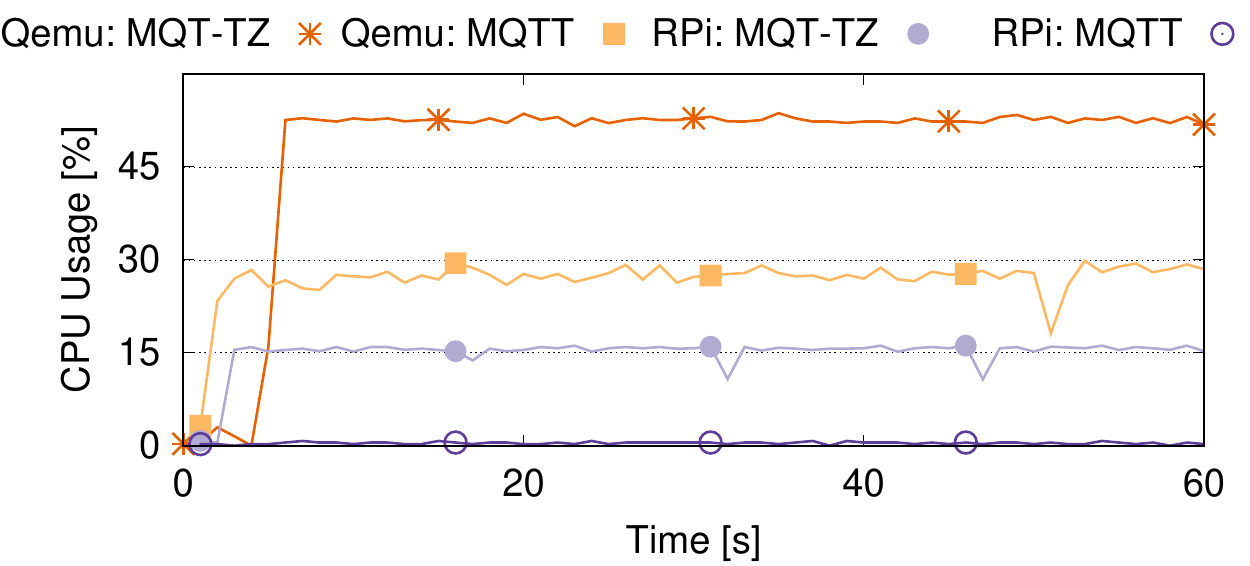}
    \caption{\sys Broker CPU usage under the load from Figure~\ref{fig:out-traffic}.\label{fig:cpu-usage}}
\end{figure}

During the experiment, we use \texttt{dstat}~\cite{dstat} to record the CPU load at the broker (see Figure~\ref{fig:cpu-usage}). 
We report the results for physical and virtualized environments, and against against a vanilla \mosquitto deployment for both scenarios.
Both in emulation and hardware, \sys adds between 10\% to 15\% to the CPU load on average, up at around 55\% usage in the worst case.
It is worth noting how these values are considerably higher in emulation than in HW, and the reduced amount of CPU ($\sim1\%$) \mosquitto requires.
Overall, this suggests low energy consumption, an important factor for deployers that intend to deploy brokers in batter-powered nodes.


\vspace{-10pt}
\section{Related Work} \label{sec:related-work}
Few attempts exist to provide secure extensions for \mqtt~\cite{shin2016security}, but none of them rely on TEEs.
Despite, TEEs (and in particular \tz) are used in several different domains, from cardiac signal processing over untrusted clouds~\cite{Segarra2019} to control-based \tz policy framework for air drones~\cite{Vijeev2019}. 
For the sake of conciseness, we shortly review secure messaging libraries on top of TEEs, as well as presenting a short survey of different usages of TEEs.

\textbf{Secure Messaging Libraries Using TEEs.}
\sys leverages the native support of \mqtt to establish TLS channels between the client and the broker.
TaLoS~\cite{talos} can establish secure TLS termination inside Intel SGX enclaves.
Deploying a complete TLS stack inside the TEE is unnecessary in our context, and it would yield a larger attack surface, as the code loaded in the \tz must be fully trusted.

PubSub-SGX~\cite{Arnautov18} is a content-based publish/subscribe framework on top of SCONE~\cite{Arnautov16},
a compilation tool chain to securely run Linux containers inside SGX enclaves.
PubSub-SGX is implemented in Python.
The notion of \emph{topic} is a first-class entity currently not supported by PubSub-SGX, making its adoption for our scenarios requiring relevant engineering efforts.
Similar drawbacks exist in~\cite{Pires2016}, a content-based routing mechanism for SGX on top of which privacy-preserving pub/sub framework can be implemented.


StreamBox-TZ~\cite{Park19} is a secure stream analytics framework for \tz, specifically targeting telemetry data.
Rather, \sys focuses on secure end-to-end packet delivery, delegating all the application-specific processing to client nodes.

\textbf{TEE-Based Applications.}
Due to the additional security guarantees and resilience to stronger adversarial models, TEEs are currently being deployed across a plethora diverse scenarios.
One important step for all the applications is the attestation protocol, \emph{e.g.}, a process used by participants to verity the integrity and validity of the trusted applications as well as the CPU executing those. 
While Intel SGX has native support for remote attestation~\cite{intel-attest}, \tz lacks clear specifications for it. 
Fides~\cite{Prunster2019} presents a ready-to-use Key attestation framework for Android's TEE, and we intend to look further into it.
A common domain of application for TEEs is web-based systems, as well as a common way for end-users to disclose personal data. 
In this context, SGX can protect the identity of the users against re-identification attacks via browser extensions and privacy proxies~\cite{Pires2018,Ben2018}.
While \sys targets scenarios where the identity of the users is not to be hidden (rather, the opposite, as in our \emph{MedTech} scenario), a broker can indeed be compromised.
Techniques such as SGX-Tor~\cite{Kim17} could be explored in the context of \tz and \sys to reduce the information shared with the brokers.



\section{Lessons Learned} \label{sec:lessons-learned}

Through the implementation, deployment and evaluation of \sys we acquired insights into several aspects of such a system.
We highlight here the most important ones.

First, the \optee framework is sufficiently mature to allow rapid development cycles to quickly test with \tz.
Trusted Applications, as our LRU cache in the TEE or the re-encryption TA, can achieve performances close to the corresponding non-secure ones. 
However, the crypto primitives they provide are slower than expected, especially when compared to other state-of-the-art libraries (\emph{i.e.}, \texttt{OpenSSL}~\cite{openssl}).
Our experimental results highlight a slowdown of up to two orders of magnitude.
We hope to mitigate these drawbacks exploiting hardware support for \tz-specific Cryptographic Hardware Accelerators such as \arm \textsc{CryptoIsland}~\cite{arm-cryptoisland}.

Second, we have shown in our evaluation that the \texttt{mosquitto} \mqtt broker has a small CPU footprint.
However, code executed in \tz can only be single-threaded, negatively affecting scalability.
While leveraging multi-threading in the broker is part of future work, this must be carefully handled.
In fact, we expect the increased CPU usage to lead to higher memory consumption, as well as higher energy requirements.
In this sense, deployers should carefully evaluate such trade-offs and decide on application-dependent requirements.

Finally, we report how the emulation accuracy for \arm processors in \textsc{Qemu} is sufficiently accurate to allow us to validate the design and implementation without having to deploy large (and potentially) expensive testbeds.
Yet, as shown in \S\ref{sec:evaluation}, the timing measurements from \textsc{Qemu} can be inaccurate, and real-hardware measurements must be planned.

\section{Conclusion and Future Work} \label{sec:conclusion}
Motivated by the lack of secure-by-design communication protocols for the edge and two real-world use-cases, we built \sys, a secure edge-based publish/subscribe middleware using \mqtt and \tz.
We report on our experiences while building and evaluating our open-source prototype against a vanilla \mqtt under real-world workloads. 
Despite the measured slowdown (up to 8$\times$ in some scenarios), our system scales and can be deployed in restricted, IoT-based settings, achieving dissemination delays in the orders of milliseconds, even when deployed in low-end devices (such as Raspberry Pi units).
In addition, the motivating use-cases described in \S\ref{sec:use-case} can benefit of the additional security guarantees provided by \sys with no additional hardware and without changes to the client's application code.

We plan to extend this work along the following directions.
Firstly, we intend to extend the \sys evaluation and compare against other topic-based publish-subscribe systems and messaging queues as well as testing in larger-scale settings with clustered \mqtt brokers.
Secondly, we intend to study the energy trade-offs of our system, a key aspect for edge deployments.
Thirdly,  we will revise \sys architecture to shield in \tz some additional components (\emph{e.g.}, ACLs, the subscription lists).
Lastly, we plan to implement a proof of concept version of \sys leveraging alternative software development kits for \tz such as \textsc{OpenEnclave} (\url{https://openenclave.io/sdk/}) and alternative \tz-enabled devices (\emph{i.e.}, \textsc{TrustBox} \url{https://scalys.com/trustbox-industrial/}).


%

\vspace{-5pt}
\section*{Acknowledgment}
This project has received funding from the European Union's Horizon 2020 research and innovation programme under grant agreement No 766733.
{\footnotesize
\bibliographystyle{plain}
\bibliography{biblio} 
}

\end{document}